\documentclass[runningheads]{llncs}

\usepackage[T1]{fontenc}
\usepackage{graphicx}
\usepackage{booktabs}
\usepackage{multirow}
\usepackage{tabularx}
\usepackage{array}
\usepackage{amsmath}
\usepackage{url}
\usepackage{CJKutf8}
\usepackage{microtype}
\usepackage{placeins}

\newcolumntype{Y}{>{\raggedright\arraybackslash}X}
\newcolumntype{R}{>{\raggedleft\arraybackslash}X}
\newcolumntype{P}[1]{>{\raggedright\arraybackslash}p{#1}}
\renewcommand{\arraystretch}{1.08}

\begin{document}
\raggedbottom

\title{Summary of the ChinaVoices Challenge 2026: Data, Tasks, Baseline, and Methods}
\titlerunning{ChinaVoices Challenge 2026}

\author{Yujie Liao\inst{1}\textsuperscript{*} \and
Bingshen Mu\inst{1}\textsuperscript{*} \and
Shuiyuan Wang\inst{1} \and
Liumeng Xue\inst{2} \and
Hexin Liu\inst{3} \and
Xian Shi\inst{4} \and
Jie Hu\inst{5} \and
Lei Xie\inst{1}\textsuperscript{**}}
\authorrunning{Y. Liao et al.}

\institute{Audio, Speech and Language Processing Group (ASLP@NPU),\\
Northwestern Polytechnical University, Xi'an, China \and
School of Intelligence Science and Technology,\\
Nanjing University \and
Nanyang Technological University \and
Independent Researcher \and
Beijing Huiting Technology Co., Ltd.}

\maketitle

\begin{center}
\footnotesize
\textsuperscript{*}These authors contributed equally.\quad
\textsuperscript{**}Corresponding author.
\end{center}

\begin{abstract}
This paper summarizes the ChinaVoices Challenge 2026, which aims to establish unified task definitions and evaluation conditions for Chinese dialect speech processing and to advance multi-dialect identification and automatic speech recognition. The challenge covers 16 dialect categories and defines two tasks: Chinese Multi-Dialect Identification and Chinese Multi-Dialect Automatic Speech Recognition (ASR). It uses approximately 320 hours of speech across the Reference Set, Open Evaluation Set, and Hidden Evaluation Set. The two tasks use the same evaluation audio, and each includes restricted-data and open-data tracks. We describe the task settings, data, evaluation metrics, and Qwen3-ASR-1.7B baseline, and analyze the leaderboard results and submitted systems. In total, 28 teams submit results, 17 provide system reports, and systems from 15 teams pass the compliance review and are included in the analysis. Most eligible systems outperform the baseline, and the official top-three order remains unchanged on the Hidden Evaluation Set for both tasks. Dialect-level results show that categories with higher identification accuracy generally have lower ASR error rates, although the tasks assess related but distinct capabilities. Leading identification systems commonly exploit dialect-discriminative acoustic representations, whereas leading ASR systems emphasize data normalization, augmentation, and auxiliary CTC objectives. These results provide practical guidance for developing and evaluating Chinese multi-dialect speech processing systems.

\keywords{Chinese dialect \and multi-dialect identification \and
multi-dialect ASR \and low-resource speech \and challenge evaluation}
\end{abstract}

\section{Introduction}
\label{sec:introduction}

Automatic speech recognition (ASR) now achieves strong performance
for standard Mandarin and is widely deployed in
applications such as voice assistants, meeting transcription, and spoken
interaction. This progress is supported by large-scale weakly
supervised learning~\cite{radford2023whisper} and self-supervised speech
representation learning~\cite{baevski2020wav2vec2}. At the same time,
multi-dialect ASR attracts growing
interest in both academia and industry. 
Real-world Chinese speech encompasses a wide range of dialect
beyond standard Mandarin. This challenge focuses on 16 selected dialect
categories, including \texttt{cantonese}, \texttt{minnan},
\texttt{kejia}, and \texttt{chaoshan}. Processing such
speech involves two related capabilities: identifying the dialect category
and accurately transcribing the spoken content. 
Although current models perform well on standard Mandarin, stable
dialect-category classification and accurate transcription remain challenging under
multi-dialect and low-resource conditions.

Research on multi-dialect identification and multi-dialect ASR explores a wide
range of approaches involving acoustic representations, data resources, and
model architectures. For multi-dialect identification, early studies primarily
use bottleneck features~\cite{zhang2017dialectrecognition} and deep acoustic
models~\cite{weninger2019accent} to learn discriminative speech
representations. More recent work exploits intermediate representations
from pretrained speech encoders, dialect speech
embeddings~\cite{chang2026embeddings}, and multi-layer feature fusion to
better distinguish acoustically similar dialects. For multi-dialect ASR,
Mandarin corpora such as AISHELL-1~\cite{bu2017aishell1},
AISHELL-3~\cite{shi2021aishell3},
AISHELL-4~\cite{fu2021aishell4}, and
WenetSpeech~\cite{zhang2022wenetspeech} provide a foundation for
training Chinese speech models, while public resources including
KeSpeech~\cite{tang2021kespeech},
WenetSpeech-Yue~\cite{li2026wenetspeechyue},
WenetSpeech-Chuan~\cite{dai2025wenetspeechchuan}, and
WenetSpeech-Wu~\cite{wang2026wenetspeechwu} extend data coverage to
\texttt{cantonese}, \texttt{sichuan}, \texttt{wuyu}, and other dialect
categories. Meanwhile, systems and models such as
FireRedASR2S~\cite{xu2026fireredasr2s},
Qwen3-ASR~\cite{shi2026qwen3asr}, and
Dolphin-CN-Dialect~\cite{meng2026dolphin} further advance the modeling
of Chinese dialect speech. Mixture-of-experts
architectures~\cite{zhou2024dialectmoe}, parameter-efficient
adaptation~\cite{bai2024adapter}, and dialect-aware representations also
provide new approaches to adapting multi-dialect models. Despite this
progress, different studies typically use different dialect inventories,
training resources, and evaluation sets, making direct comparison under a common
evaluation setting difficult.

Chinese dialect speech processing still lacks a unified, public, and
reproducible evaluation platform. A major limitation of existing research is the lack of a public and standardized evaluation set covering a broad range of Chinese dialects. Existing publicly available evaluation sets cover only a subset of dialects, leaving many regional varieties without a common benchmark for fair and reproducible evaluation.
These limitations also hinder 
continued progress in Chinese dialect speech technologies. Therefore, a unified 
Chinese multi-dialect platform is needed to evaluate both multi-dialect identification 
and multi-dialect ASR, with independent verification procedures to ensure fair evaluation and reliable results.

To address this need, the ChinaVoices Challenge 2026 is organized in
conjunction with NCMMSC 2026. The challenge defines two tasks:
multi-dialect identification and multi-dialect ASR. It establishes
evaluation data, metrics, and competition procedures
for 16 dialect categories. For each dialect category, it provides
approximately 3 hours in the Reference Set, 7 hours in the Open Evaluation
Set, and 10 hours in the Hidden Evaluation Set. The three sets are
speaker-disjoint, and all ground-truth transcripts are manually produced and
quality-checked. The two tasks use the same evaluation audio and measure
dialect-category classification and content transcription using
macro-averaged accuracy (ACC) and macro-averaged character error rate (CER),
respectively. Each task contains a restricted-data track and an open-data
track, yielding four leaderboards in total. The restricted-data tracks
explicitly constrain the data resources that may be used and
determine the official rankings, whereas the open-data tracks permit broader
legally usable data to explore the performance ceiling under the same
evaluation setting. This paper also introduces the baseline and analyzes the
participant systems in terms of overall results, dialect-level performance,
error patterns, and system design. The complete challenge information is
available on the official
website.\footnote{\url{https://aslp-lab.github.io/ChinaVoices-Challenge/}}

\section{Challenge Tasks and Tracks}
\label{sec:tasks}

The ChinaVoices Challenge comprises multi-dialect identification and
multi-dialect ASR. The tasks share the same evaluation audio and 16 dialect
categories but differ in output and metric. Each task contains a
restricted-data track and an open-data track, yielding four leaderboards in
total. Teams may enter either task or both and may use separate task models or
one unified model.

\subsection{Multi-Dialect Identification}
\label{sec:task-did}

Given a single utterance, the multi-dialect identification task requires one
of 16 dialect labels and no transcript. The label inventory consists of \texttt{anhui},
\texttt{cantonese}, \texttt{changsha}, \texttt{chaoshan},
\texttt{dongbei}, \texttt{henan}, \texttt{kejia}, \texttt{minnan},
\texttt{nanchang}, \texttt{nanjing}, \texttt{shan1xi},
\texttt{shan3xi}, \texttt{shandong}, \texttt{sichuan},
\texttt{wuhan}, and \texttt{wuyu}. These official labels are used
consistently throughout the paper. In particular, \texttt{shan1xi} and
\texttt{shan3xi} denote Shanxi and Shaanxi, respectively.

Let class \(k\) contain \(N_k\) evaluation utterances, of which
\(C_k\) are classified correctly. Its accuracy is
\begin{equation}
  \mathrm{ACC}_k = \frac{C_k}{N_k}\times 100\%.
\end{equation}
The official metric is the macro-averaged ACC over the 16 dialect
categories:
\begin{equation}
  \mathrm{ACC}_{\mathrm{macro}}
  = \frac{1}{K}\sum_{k=1}^{K}\mathrm{ACC}_k,
  \qquad K=16.
\end{equation}
Macro-averaging gives every dialect category equal weight; higher values
indicate better multi-dialect identification.

\subsection{Multi-Dialect ASR}
\label{sec:task-asr}

The multi-dialect ASR task requires a system to transcribe an input
utterance into a character sequence. One principal model or unified framework
must process all 16 dialect categories. 

The evaluation metric is character error rate. For dialect \(k\), let its
ground-truth transcripts contain \(M_k\) characters, and let \(S_k\), \(D_k\),
and \(I_k\) denote the numbers of substitutions, deletions, and insertions
after optimal edit alignment. The dialect-level CER is
\begin{equation}
  \mathrm{CER}_k
  = \frac{S_k+D_k+I_k}{M_k}\times 100\%.
\end{equation}
The official metric is the macro-averaged CER across all dialect categories:
\begin{equation}
  \mathrm{CER}_{\mathrm{macro}}
  = \frac{1}{K}\sum_{k=1}^{K}\mathrm{CER}_k,
  \qquad K=16.
\end{equation}
This macro-averaged metric again weights dialect categories equally; lower
values indicate better transcription. The official evaluation script defines
text normalization and metric computation.

\subsection{Tracks and Leaderboards}
\label{sec:data-tracks}

The restricted-data track permits organizer-provided data, approved public
corpora, public pretrained models, and reproducible training resources; it
determines official rankings and awards. The open-data track additionally
allows legally usable internal or self-collected data and is reported for
reference only. Teams disclose the principal data, models, augmentation, and
post-processing used in each system. Both tracks require offline inference
with one principal model or unified framework rather than a cascade of
independent dialect-specific models. The leaderboards report scores on the
Open Evaluation Set, while leading restricted-data systems undergo compliance,
reproduction, and independent evaluation on the unreleased Hidden Evaluation
Set. Official placements follow this review rather than the Open Evaluation
Set scores alone.

\section{Challenge Data and Baseline System}
\label{sec:data-baseline}

\subsection{Challenge Data}
\label{sec:data}

The challenge data cover all 16 dialect categories. Every utterance has a manually produced,
quality-checked ground-truth transcript.
The Reference Set provides audio, labels, and transcripts for approximately
3 hours per dialect. The Open Evaluation Set releases only audio for approximately
7 hours per dialect and determines public leaderboard scores. The unreleased
Hidden Evaluation Set contains approximately 10 hours per dialect and is used
to verify leading restricted-data systems independently. The three sets are
strictly speaker-disjoint, preventing systems from exploiting voices observed
during training. Neither the Open Evaluation Set nor the Hidden Evaluation
Set may be used for training, fine-tuning, pseudo-labeling, or
evaluation-targeted data construction.

The restricted-data track also permits representative public resources such
as AISHELL-1,
AISHELL-3,
AISHELL-4,
WenetSpeech,
KeSpeech,
WenetSpeech-Yue,
WenetSpeech-Chuan, and
WenetSpeech-Wu, together with listed corpora for
\texttt{kejia}, \texttt{minnan}, and other covered varieties.
Resources may be combined subject to their licenses, but provenance and
processing must remain public and traceable. The challenge website provides
the complete resource list.

The restricted-data track controls resource availability rather than
forcing every team to use the same mixture. Teams can select corpora according
to dialect coverage and licensing, so system reports remain necessary for
interpreting differences in training composition while retaining public
reproducibility.

\subsection{Unified Baseline}
\label{sec:baseline}

The baseline, based on the public pretrained model Qwen3-ASR-1.7B,
uses supervised fine-tuning through
\texttt{ms-swift} \cite{zhao2025swift}. It formulates dialect-label
prediction and transcription as a unified conditional generation task: the
audio encoder and audio--text aligner produce acoustic representations, and
the language model generates the dialect label followed by the transcript,
separated by the \texttt{\textless asr\_text\textgreater} token. For example, a target sequence is formatted as follows: \begin{quote} \small\ttfamily language Chinese chaoshan\textless asr\_text\textgreater \begin{CJK*}{UTF8}{gbsn}今天天气不错\end{CJK*} \end{quote} The generated dialect label is used for multi-dialect identification, while the transcript following the separator is used for multi-dialect ASR.
Thus, a
single generation process and checkpoint support both tasks. Training
combines approved public dialect corpora with a speaker-disjoint internal
training/development split of the Reference Set. LoRA
\cite{hu2022lora} adapts the language-model component, while the audio encoder
and aligner remain frozen. On the Open Evaluation Set, the resulting
checkpoint achieves 53.62\% macro-averaged ACC for multi-dialect
identification and 18.10\% macro-averaged CER for multi-dialect ASR. Complete data manifests,
training configurations, and reproduction scripts are available in the
challenge repository.\footnote{\url{https://github.com/ASLP-lab/ChinaVoices-Challenge}}

\section{Results and Analysis}
\label{sec:results}

The challenge receives leaderboard entries from 28 teams, and 17 of them
also submit system reports. After review of participant eligibility, data
usage, system descriptions, and result validity, systems from 15 eligible
teams meet the challenge requirements and are included in the result and
system analyses. We first report the four leaderboards and then analyze
dialect-level performance, error patterns, and system designs.

\subsection{Overall Results}
\label{sec:overall-results}

Tables~\ref{tab:did-leaderboard} and \ref{tab:asr-leaderboard} report the full
leaderboards for multi-dialect identification and multi-dialect ASR,
respectively. The restricted-data leaderboards determine the official
placements, whereas the open-data leaderboards report only Open Evaluation
Set results. Official ranks are marked only for the reviewed top three in each
restricted-data leaderboard. The baseline serves as a common reference and is
not counted as a participating system.


\begin{table}[htb!]
\caption{Multi-dialect identification leaderboard.}
\label{tab:did-leaderboard}
\centering
\small
\setlength{\tabcolsep}{4.5pt}
\renewcommand{\arraystretch}{1.05}

\begin{tabularx}{\textwidth}
{@{}l c >{\raggedright\arraybackslash}X c c@{}}
\toprule
Track & Rank & Team &
\multicolumn{1}{c}{\shortstack{Open Eval.\\ACC (\%)}} &
\multicolumn{1}{c}{\shortstack{Hidden Eval.\\ACC (\%)}} \\
\midrule

\multirow[t]{9}{*}{Restricted-data}
 & 1  & scy919           & 83.19 & 79.75 \\
 & 2  & Optima           & 78.47 & 73.08 \\
 & 3  & zenava.ai        & 73.42 & 69.54 \\
 & -- & 4pd              & 67.57 & \multicolumn{1}{c}{--} \\
 & -- & HuofanTeam       & 67.50 & \multicolumn{1}{c}{--} \\
 & -- & voxmindlabs      & 65.95 & \multicolumn{1}{c}{--} \\
 & -- & kairot           & 56.53 & \multicolumn{1}{c}{--} \\
 & -- & Mlslabs          & 56.13 & \multicolumn{1}{c}{--} \\
 & -- & Unified baseline & 53.62 & \multicolumn{1}{c}{--} \\
\midrule

\multirow[t]{5}{*}{Open-data}  
 & -- & Optima           & 80.65 & \multicolumn{1}{c}{--} \\
 & -- & scy919           & 80.03 & \multicolumn{1}{c}{--} \\
 & -- & zenava.ai        & 73.42 & \multicolumn{1}{c}{--} \\
 & -- & voxmindlabs      & 65.95 & \multicolumn{1}{c}{--} \\
 & -- & Unified baseline & 53.62 & \multicolumn{1}{c}{--} \\
\bottomrule
\end{tabularx}

\vspace{2pt}
\parbox{\textwidth}{\footnotesize
A dash indicates the absence of an official rank or a released Hidden Evaluation Set
score. Open-data systems receive no official ranks.}
\end{table}

\begin{table}[htb!]
\caption{Multi-dialect ASR leaderboard.}
\label{tab:asr-leaderboard}
\centering
\small
\setlength{\tabcolsep}{4.5pt}
\renewcommand{\arraystretch}{1.05}

\begin{tabularx}{\textwidth}
{@{}l c >{\raggedright\arraybackslash}X c c@{}}
\toprule
Track & Rank & Team &
\multicolumn{1}{c}{\shortstack{Open Eval.\\CER (\%)}} &
\multicolumn{1}{c}{\shortstack{Hidden Eval.\\CER (\%)}} \\
\midrule

\multirow[t]{12}{*}{Restricted-data}
 & 1  & TeleASR              & 11.08 & 10.70 \\
 & 2  & Mlslabs              & 11.22 & 11.18 \\
 & 3  & Kyoto-Tsinghua Team & 11.75 & 11.68 \\
 & -- & 4pd                  & 13.42 & \multicolumn{1}{c}{--} \\
 & -- & tencent-asr          & 13.48 & \multicolumn{1}{c}{--} \\
 & -- & CyberTech            & 13.92 & \multicolumn{1}{c}{--} \\
 & -- & Optima               & 14.49 & \multicolumn{1}{c}{--} \\
 & -- & voxmindlabs          & 15.67 & \multicolumn{1}{c}{--} \\
 & -- & nisseslab            & 16.36 & \multicolumn{1}{c}{--} \\
 & -- & Unified baseline     & 18.10 & \multicolumn{1}{c}{--} \\
 & -- & HuofanTeam           & 18.18 & \multicolumn{1}{c}{--} \\
 & -- & Nanqiang Beitiao     & 28.58 & \multicolumn{1}{c}{--} \\
\midrule

\multirow[t]{6}{*}{Open-data}
 & -- & TeleASR          &  7.42 & \multicolumn{1}{c}{--} \\
 & -- & USTC-U2Lab       &  7.45 & \multicolumn{1}{c}{--} \\
 & -- & tencent-asr      & 14.38 & \multicolumn{1}{c}{--} \\
 & -- & Optima           & 14.50 & \multicolumn{1}{c}{--} \\
 & -- & voxmindlabs      & 15.68 & \multicolumn{1}{c}{--} \\
 & -- & Unified baseline & 18.10 & \multicolumn{1}{c}{--} \\
\bottomrule
\end{tabularx}

\vspace{2pt}
\parbox{\textwidth}{\footnotesize
A dash indicates the absence of an official rank or a released Hidden Evaluation Set
score. Open-data systems receive no official ranks.}
\end{table}

In the restricted-data multi-dialect identification leaderboard, all eight
eligible systems outperform the 53.62\% baseline. The
top-ranked scy919 system reaches 83.19\% ACC, an absolute improvement of
29.57 percentage points over the baseline. Its lead over Optima is 4.72
percentage points, and the gap between Optima and zenava.ai is 5.06
percentage points. The best and worst eligible systems differ by 27.07
percentage points. Thus, despite the
substantial improvement over the common baseline, the systems still exhibit
considerable performance variation, and the leading ranks form a clearly
separated gradient.

In the restricted-data ASR leaderboard, 9 of the 11 eligible systems
outperform the 18.10\% CER baseline. TeleASR reduces CER to 11.08\%, an
absolute decrease of 7.02 percentage points and a relative error reduction
of 38.80\%. The top-three CER values span only 0.67 percentage points: the
first--second and second--third gaps are 0.13 and 0.53 percentage points,
respectively. Measured by the
absolute gaps in their respective official metrics, the leading
restricted-data ASR systems are more closely grouped than the leading
multi-dialect identification systems, although the full ASR leaderboard still
shows a substantial performance range.

The open-data tracks receive fewer eligible submissions. Their results are
therefore more suitable for observing the performance attainable by complete
systems under broader resource rules than for estimating the average
benefit of external data. The best open-data dialect-identification ACC is
80.65\%, below the restricted-data best of 83.19\%, whereas the best
open-data ASR CER is 7.42\%, 3.66 percentage points below the
restricted-data best.
Among teams entering both data tracks, Optima's dialect-identification ACC
increases by 2.18 percentage points, whereas scy919's decreases by 3.17
percentage points. For ASR, TeleASR's CER decreases by 3.66 percentage
points, whereas those of tencent-asr, Optima, and voxmindlabs increase by
0.90, 0.01, and 0.01 percentage points, respectively. Thus, submissions
to the open-data track do not consistently outperform their restricted-data
counterparts. On the Hidden Evaluation Set, the official
top-three order remains identical to that on the Open Evaluation Set for
both tasks. ACC decreases by 3.44, 5.39, and 3.88 percentage points for
the three
multi-dialect identification systems, while the CER changes of the three ASR
systems are all within 0.38 percentage points. Because the two evaluation
sets are not
difficulty-calibrated and hidden scores are released only for the top three,
these observations establish stability at the head of the leaderboards but
do not support a comparison of overall generalization between the two tasks.

\subsection{Dialect-Level Metrics and Error Analysis}
\label{sec:dialect-analysis}

\begin{table}[!htb]
\caption{Dialect-level ACC (\%) by official challenge label for the
top-three restricted-data systems on the Open Evaluation Set.}
\label{tab:did-dialect-results}
\centering
\footnotesize
\setlength{\tabcolsep}{7pt}
\renewcommand{\arraystretch}{1.02}
\begin{tabular}{@{}lcccc@{}}
\toprule
Challenge label &
\shortstack{scy919} &
\shortstack{Optima} &
\shortstack{zenava.ai} &
Mean ACC \\
\midrule
\texttt{anhui}     & 83.81 & 95.88 & 76.45 & 85.38 \\
\texttt{cantonese} & 97.10 & 99.98 & 98.77 & 98.62 \\
\texttt{changsha}  & 74.51 & 79.08 & 70.61 & 74.73 \\
\texttt{chaoshan}  & 68.53 & 72.44 & 60.30 & 67.09 \\
\texttt{dongbei}   & 99.39 & 99.69 & 99.35 & 99.48 \\
\texttt{henan}     & 88.11 & 72.17 & 85.37 & 81.88 \\
\texttt{kejia}     & 61.90 & 23.57 & 44.27 & 43.25 \\
\texttt{minnan}    & 79.14 & 84.93 & 40.33 & 68.14 \\
\texttt{nanchang}  & 77.98 & 61.44 & 64.35 & 67.92 \\
\texttt{nanjing}   & 86.17 & 70.68 & 44.40 & 67.08 \\
\texttt{shan1xi}   & 77.34 & 49.28 & 71.36 & 65.99 \\
\texttt{shan3xi}   & 95.97 & 98.18 & 92.53 & 95.56 \\
\texttt{shandong}  & 86.99 & 95.17 & 77.79 & 86.65 \\
\texttt{sichuan}   & 85.36 & 97.71 & 84.99 & 89.35 \\
\texttt{wuhan}     & 77.10 & 70.52 & 77.58 & 75.07 \\
\texttt{wuyu}      & 91.68 & 84.83 & 86.19 & 87.57 \\
\bottomrule
\end{tabular}
\end{table}

The following analysis reports the two tasks separately. In both
dialect-level tables, the systems are ordered by official rank and the mean
column is computed from their unrounded scores.

\subsubsection{Multi-Dialect Identification Difficulty and Confusions}

Table~\ref{tab:did-dialect-results} reports the Open Evaluation Set ACC of the
official top-three restricted-data multi-dialect identification systems.

\texttt{dongbei}, \texttt{cantonese}, and \texttt{shan3xi} are the easiest
dialect categories to identify, with mean ACC values of 99.48\%, 98.62\%, and
95.56\%, respectively. The \texttt{kejia} dialect category is the most
difficult, with a mean ACC
of only 43.25\%. The \texttt{shan1xi}, \texttt{nanjing}, \texttt{chaoshan},
and \texttt{nanchang} dialect categories also have mean ACC below 68.00\%. These
results indicate that the performance bottleneck in multi-dialect
identification is concentrated in a small set of dialect categories rather
than distributed uniformly across all categories. At the same time, the three
systems differ substantially on \texttt{minnan}, \texttt{nanjing}, and
\texttt{kejia}, suggesting that some difficult dialect categories still
offer room for complementary system behavior. Low mean ACC and large
between-system variation should therefore be interpreted separately as
shared difficulty and system-specific differences.

\begin{figure}[htb!]
\centering
\includegraphics[width=0.84\textwidth]{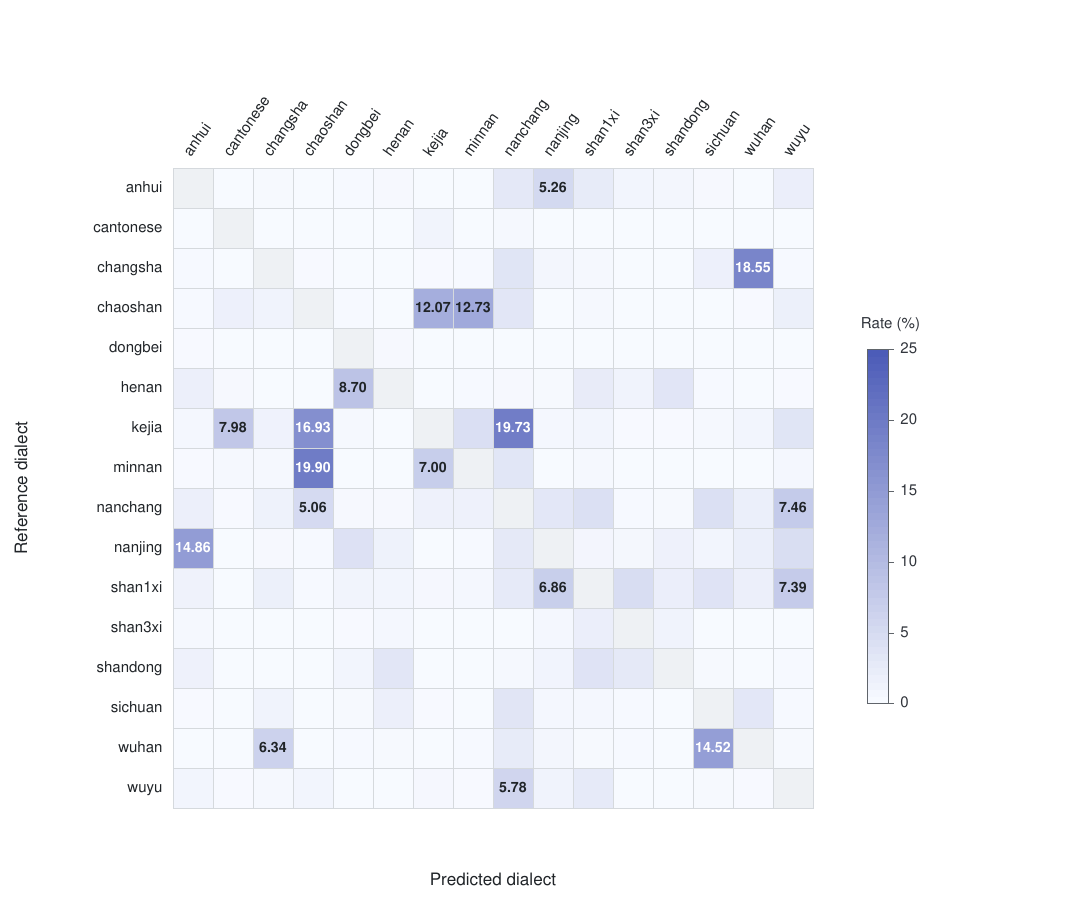}
\caption{Mean row-normalized confusion matrix (\%) of the official top-three
multi-dialect identification systems. The diagonal is masked, and only
off-diagonal values of at least 5.00\% are annotated.}
\label{fig:did-confusion}
\end{figure}

Figure~\ref{fig:did-confusion} presents the mean row-normalized confusion
matrix of the official top-three systems. The main error flows occur between
\texttt{minnan} and \texttt{chaoshan} and between \texttt{kejia} and
\texttt{chaoshan}. The \texttt{minnan} dialect category is classified as
\texttt{chaoshan} in 19.90\% of cases on average, compared with 12.73\% in the
reverse direction. The \texttt{kejia} dialect category is classified as
\texttt{chaoshan} in 16.93\% of cases, compared with 12.07\% in reverse.
\texttt{kejia} is also classified as \texttt{nanchang} in 19.73\% of cases,
showing that its errors are not concentrated on a single neighboring dialect
category.
Several category pairs exhibit marked directional asymmetry. For example,
\texttt{changsha} is classified as \texttt{wuhan} in 18.55\% of cases,
whereas the reverse rate is only 6.34\%; \texttt{nanjing} is classified as
\texttt{anhui} at 14.86\%, compared with 5.26\% in reverse. These asymmetric
patterns suggest that the errors may reflect
not only dialect similarity but also within-category variability, sample
coverage, and the learned decision boundaries.

From a system-design perspective, improving the difficult dialects is
unlikely to require only a uniform increase in training samples. The observed
confusion directions suggest constructing hard negative examples among
similar dialects, exploiting intermediate or multi-layer acoustic
representations, and strengthening category-boundary modeling.

\subsubsection{Multi-Dialect ASR Difficulty and Error Composition}

\begin{table}[!htb]
\caption{ASR CER (\%) by official challenge label for the top-three
restricted-data systems on the Open Evaluation Set.}
\label{tab:asr-dialect-results}
\centering
\footnotesize
\setlength{\tabcolsep}{7pt}
\renewcommand{\arraystretch}{1.02}
\begin{tabular}{@{}lcccc@{}}
\toprule
Challenge label &
\shortstack{TeleASR} &
\shortstack{Mlslabs} &
\shortstack{Kyoto-Tsinghua\\Team} &
Mean CER \\
\midrule
\texttt{anhui}     & 7.09  & 7.21  & 7.65  & 7.32 \\
\texttt{cantonese} & 7.60  & 7.09  & 6.89  & 7.19 \\
\texttt{changsha}  & 10.32 & 8.09  & 8.58  & 9.00 \\
\texttt{chaoshan}  & 18.39 & 23.99 & 23.28 & 21.89 \\
\texttt{dongbei}   & 5.70  & 4.92  & 5.10  & 5.24 \\
\texttt{henan}     & 7.82  & 6.94  & 6.95  & 7.24 \\
\texttt{kejia}     & 26.98 & 29.28 & 36.77 & 31.01 \\
\texttt{minnan}    & 17.17 & 13.50 & 14.45 & 15.04 \\
\texttt{nanchang}  & 12.51 & 12.54 & 13.61 & 12.89 \\
\texttt{nanjing}   & 7.67  & 7.35  & 7.05  & 7.36 \\
\texttt{shan1xi}   & 17.43 & 16.26 & 19.44 & 17.71 \\
\texttt{shan3xi}   & 5.11  & 4.73  & 5.03  & 4.95 \\
\texttt{shandong}  & 6.11  & 5.61  & 6.17  & 5.96 \\
\texttt{sichuan}   & 7.25  & 5.92  & 5.72  & 6.30 \\
\texttt{wuhan}     & 5.66  & 4.91  & 4.71  & 5.09 \\
\texttt{wuyu}      & 14.48 & 21.08 & 16.55 & 17.37 \\
\bottomrule
\end{tabular}
\end{table}

Table~\ref{tab:asr-dialect-results} reports the Open Evaluation Set CER of the
official top-three restricted-data ASR systems.

\texttt{kejia} is also the most difficult dialect category for ASR, with a
mean CER of
31.01\% across the three systems. The \texttt{chaoshan}, \texttt{shan1xi},
\texttt{wuyu}, and \texttt{minnan} dialect categories form a second difficult
group,
with mean CER values of 21.89\%, 17.71\%, 17.37\%, and 15.04\%,
respectively. The \texttt{shan3xi}, \texttt{wuhan}, and \texttt{dongbei}
dialect categories are relatively easy, with mean CER values of 4.95\%, 5.09\%, and
5.24\%. Dialect-level
strengths are not consistent across systems: TeleASR, Mlslabs, and the
Kyoto-Tsinghua Team obtain the lowest CER among the three systems on 5, 7,
and 4 dialects, respectively. Their similar overall leaderboard scores but
different dialect-level optima indicate that the training pipelines affect
individual dialects differently.

We further analyze substitution, deletion, and insertion errors (S/D/I).
Table~\ref{tab:asr-sdi} reports their micro-averaged rates for the official
top-three systems. These statistics use all ground-truth characters as the
denominator and therefore differ slightly from the 16-dialect macro-averaged
CER used for the leaderboard.

\begin{table}[htb!]
\caption{Micro-averaged ASR error rates (\%) of the official top-three
restricted-data systems on the Open Evaluation Set.}
\label{tab:asr-sdi}
\centering
\small
\setlength{\tabcolsep}{7pt}
\begin{tabular}{@{}lcccc@{}}
\toprule
System & Substitution & Deletion & Insertion &
\shortstack{Micro-averaged\\CER} \\
\midrule
TeleASR & 9.74 & 0.81 & 0.61 & 11.17 \\
Mlslabs & 9.38 & 1.29 & 0.63 & 11.30 \\
Kyoto-Tsinghua Team & 9.79 & 1.41 & 0.62 & 11.82 \\
\bottomrule
\end{tabular}
\vspace{2pt}

\parbox{0.90\textwidth}{\footnotesize Micro-averaged CER is computed from unrounded
error counts; displayed components may not sum exactly to the reported total
because of rounding.}
\end{table}

Substitutions account for 82.80--87.30\% of all edit errors in the three
systems and are the dominant error type. Insertions account for only
5.30--5.60\%, and the overall insertion rates cluster between 0.61\% and
0.63\%; they are therefore not the primary source of the ranking
differences. Deletion rates vary more across systems, with TeleASR obtaining
the lowest rate. Although TeleASR has a higher substitution rate than
Mlslabs, its lower deletion rate yields a slightly lower overall CER, showing
that similar total CER values can arise from different error compositions.

The high CER of difficult dialects is driven primarily by accumulated
substitutions. Substitution rates range from 24.22\% to 31.72\% for
\texttt{kejia}, from 16.04\% to 20.64\% for \texttt{chaoshan}, and from
14.79\% to 17.50\% for \texttt{shan1xi}. Deletions also exhibit notable
system variation for \texttt{wuyu} and \texttt{kejia}, whose deletion rates
range from 0.77\% to 3.53\% and from 1.34\% to 3.46\%, respectively. These
results suggest that future systems should prioritize
lexical and pronunciation coverage for difficult dialects while combining
difficult-dialect sampling, auxiliary acoustic--text alignment, and
cross-corpus text normalization.

\subsubsection{Relationship Between the Two Tasks}

Using the mean dialect-level scores of the official top-three systems in each
task, we examine whether dialect-identification difficulty is associated with
ASR difficulty. The official challenge labels are used throughout this
analysis. Across the 16 dialect categories, dialect-identification ACC and
dialect-level CER show a strong
negative association (Pearson's \(r=-0.76\) and Spearman's
\(\rho=-0.74\)). In general, dialect categories with higher
dialect-identification ACC tend to have lower dialect-level CER. The
\texttt{kejia}, \texttt{chaoshan}, \texttt{minnan}, and
\texttt{shan1xi} dialect categories are difficult in both tasks, whereas
\texttt{dongbei} and \texttt{shan3xi} are relatively easy in both. This
shared pattern suggests that performance in the two tasks may be partly
affected by common factors, such as acoustic separability, training-data
coverage, and domain mismatch. Because the analysis uses dialect-level
averages from different systems, however, the observed association is
descriptive and does not imply a causal relationship between the two tasks.

Several dialect categories deviate from the overall trend. The
\texttt{wuyu} dialect category
achieves a high mean ACC of 87.57\% but also a relatively high mean CER of
17.37\%, indicating that it is comparatively easy to identify but difficult
to transcribe. In contrast, \texttt{nanjing} has a mean ACC of only 67.08\%
but a mean CER of 7.36\%; \texttt{wuhan} similarly combines moderate ACC
(75.07\%) with low CER (5.09\%). These exceptions show that the two tasks
capture related but distinct aspects of dialect speech: salient
dialect-identity cues do not necessarily provide sufficient lexical and
pronunciation knowledge for transcription, while confusion between dialect
categories does not always imply poor ASR performance. Multi-dialect
identification information may therefore serve as a complementary signal for
ASR, but it
cannot replace direct evaluation of transcription performance.

\subsection{System Analysis}
\label{sec:design-patterns}

\subsubsection{Multi-Dialect Identification Systems}

The top three restricted-data systems follow three different
multi-dialect identification routes. scy919 constructs a dual encoder based on FireRedLID
and Dolphin-CN-Dialect to extract complementary intermediate-layer
features. The extracted features are then fed into a TabPFN back-end
classifier \cite{hollmann2025tabpfn} to predict the dialect category. Optima primarily uses multi-layer
FireRedLID features and combines several training objectives, including
character-level CTC, acoustic-unit-level CTC \cite{graves2006ctc}, and
auxiliary dialect-family classification. It then fuses representations from
multiple levels and acoustic views. zenava.ai instead adapts
Qwen3-Omni-30B-A3B \cite{xu2025qwen3omni} with LoRA and
formulates multi-dialect identification as generative category coding.

The recurring properties among these high-ranking systems are
dialect-specific representations, intermediate or multi-layer features, and
complementary decision mechanisms rather than model scale alone. Because the
results come from complete systems developed by different teams, the
comparison indicates recurring design choices but cannot estimate the
independent contribution of any component.

\subsubsection{Multi-Dialect ASR Systems}

All three leading restricted-data systems use FireRedASR2-AED
as their foundation model but adopt different
training pipelines. TeleASR emphasizes cross-corpus transcription
normalization and alignment, extends the information available to short
utterances through context concatenation, and uses speed perturbation
\cite{ko2015audioaugmentation}, noise injection, and other augmentation
methods. Mlslabs applies staged freezing and unfreezing, joint AED--CTC
constraints \cite{kim2017jointctcattention}, and oversampling
of difficult dialects. The Kyoto-Tsinghua Team uses layer-wise learning
rates and an auxiliary CTC objective and employs SeedVC
\cite{liu2024seedvc} to increase speaker and acoustic diversity.

These systems improve the foundation model at the levels of data coverage
and acoustic--text alignment. TeleASR focuses on transcription
normalization, context expansion, and augmentation, whereas Mlslabs and the
Kyoto-Tsinghua Team use auxiliary CTC objectives to strengthen sequence
alignment. Other high-ranking systems also use vocabulary expansion,
unknown-symbol suppression, and beam search to improve output consistency.
Overall, competitive ASR systems tend to optimize data processing, training
objectives, and decoding jointly rather than merely replacing the foundation
model with a stronger one.

\subsubsection{Foundation Model Choice and Final Performance}

FireRedASR2-AED is concentrated near the top of the
restricted-data ASR leaderboard. The five systems using this model obtain CER values from
11.08\% to 13.92\%, with a median of 11.75\%, and five of the top six
restricted-data systems use this foundation model. This distribution
indicates that a strong foundation model can provide a high performance
starting point. Nevertheless, the best and worst results with the same model
still differ by 2.84 percentage points in CER, and the dialect-level optima of the
top three systems are distributed differently. Data normalization,
difficult-dialect sampling, auxiliary alignment, augmentation, and decoding
therefore remain substantial sources of complete-system variation.

Restricted-data ASR systems based on Qwen3-ASR span
13.48\%--16.36\% CER and likewise exhibit considerable variation. For
multi-dialect identification, frozen dual encoders, deeper multi-objective
adaptation, and LoRA all have high-ranking representatives. The current
results do not show a monotonic relationship between model size or the
fraction of updated parameters and leaderboard rank. A more cautious
conclusion is that the foundation model influences the performance starting
point but does not determine the final result by itself; final performance
also depends on task-matched representations, data processing, training
objectives, and inference and decoding strategies. 



\section{Conclusion}
\label{sec:conclusion}

This paper presents the ChinaVoices Challenge 2026, a unified platform covering 16 dialect categories and two tasks: multi-dialect identification and multi-dialect ASR. The two tasks share the same evaluation audio, which is organized into three speaker-disjoint sets: the Reference Set, Open Evaluation Set, and Hidden Evaluation Set. Each task contains a restricted-data track and an open-data track, supporting both official evaluation under controlled resource rules and performance exploration with broader data resources. The unified baseline based on Qwen3-ASR-1.7B achieves 53.62\% macro-averaged ACC and 18.10\% macro-averaged CER on the Open Evaluation Set, whereas the best restricted-data systems reach 83.19\% ACC and 11.08\% CER. The official top-three order remains unchanged on the Hidden Evaluation Set in both tasks. Dialect-level analysis reveals a general association between dialect-identification ACC and dialect-level CER, while several exceptions show that the two tasks measure related but distinct modeling capabilities. High-ranking systems adopt techniques such as dialect-aware or multi-layer acoustic representations, data processing and augmentation, difficult-dialect sampling, and auxiliary CTC objectives. The results also indicate that the foundation model influences the performance starting point but does not determine the final outcome by itself. Same-team submissions to the two data tracks exhibit mixed performance changes, so the current results do not isolate the benefit of broader training resources. ChinaVoices Challenge 2026 provides a common task and evaluation platform for Chinese multi-dialect speech processing. Future editions could broaden real-world scenario coverage, refine dialect transcription guidelines, standardize system reporting, and provide more comprehensive dialect-level performance and error statistics.

\bibliographystyle{splncs04}
\bibliography{references}

\end{document}